\documentclass[conference]{IEEEtran}
\IEEEoverridecommandlockouts
\usepackage{algorithm2e}
\usepackage[binary-units]{siunitx}
\DeclareSIUnit{\dBm}{\deci\belmilliwatt}
\usepackage{tabularx}
\usepackage{amsmath,amssymb,amsfonts}
\usepackage{color, colortbl}
\DeclareSIUnit{\bps}{bps}
\usepackage{graphicx}
\usepackage{cite}
\usepackage{flushend}

\begin{document}
\title{Dual-Beam Intelligent Reflecting Surface for Millimeter and THz Communications}

\author{\IEEEauthorblockN{Wei Jiang\IEEEauthorrefmark{1}\IEEEauthorrefmark{2} and Hans D. Schotten\IEEEauthorrefmark{2}\IEEEauthorrefmark{1}}
\IEEEauthorblockA{\IEEEauthorrefmark{1}German Research Center for Artificial Intelligence (DFKI), Trippstadter Str. 122,  Kaiserslautern, 67663 Germany\\
  }
\IEEEauthorblockA{\IEEEauthorrefmark{2}Technische Universit\"at (TU) Kaiserslautern, Building 11, Paul-Ehrlich Street, Kaiserslautern, 67663 Germany\\
 }
\thanks{This work was supported by the German Federal Ministry of Education and Research (BMBF) through the \emph{Open6G-Hub} project (Grant no.  \emph{16KISK003K}).
}
}

\maketitle

\begin{abstract}
Intelligent reflecting surface (IRS) is a cost-efficient technique to improve power efficiency and spectral efficiency. However, IRS-aided multi-antenna transmission needs to jointly optimize the passive and active beamforming, imposing a high computational burden and high latency due to its iterative optimization process. Making use of hybrid analog-digital beamforming in high-frequency transmission systems, a novel technique, coined dual-beam IRS, is proposed in this paper. The key idea is to form a pair of beams towards the IRS and user, respectively. Then, the optimization of passive and active beamforming can be decoupled, resulting in  a simplified system design. Simulation results corroborate that it achieves a good balance between the cell-edge and cell-center performance. Compared with the performance bound, the gap is moderate, but it remarkably outperforms other sub-optimal schemes.

\end{abstract}

%
\IEEEpeerreviewmaketitle

\section{Introduction}
Since the commercial deployment of the fifth-generation (5G) technology, academia and industry worldwide have initiated many research programs on the sixth generation (6G) technology \cite{Ref_jiang2021kickoff}.  The ITU-T focus group \textit{Technologies for Network 2030} envisioned that 6G will support disruptive applications - holographic-type communications, extended reality, digital twin, to name a few -  imposing extreme performance requirements, such as a peak data rate of 1 terabits-per-second (Tbps) and missive connection with a density of $10^7$ devices per \si{\kilo\meter^2} \cite{Ref_jiang2021road}. As of today, the total bandwidth assigned for 5G millimeter-wave (mmWave) communications by the ITU-R at the World Radiocommunication Conference 2019 amounts to \SI{13.5}{\giga\hertz}. Using such bandwidth, a data rate of \SI{1}{\tera\bps} can only be achieved under a spectral efficiency approaching \SI{100}{\bps/\hertz}, which requires symbol fidelity that is infeasible using currently known transmission techniques or transceiver components. Consequently, 6G needs to exploit the massively abundant spectrum at higher frequencies, including the mmWave band from \SIrange{30}{300}{\giga\hertz}, and the terahertz (THz) band from \SIrange{0.1}{3}{\tera\hertz} \cite{Ref_jiang2022initial}. At the WRC-19, the ITU-R has already identified the spectrum between \SI{275}{\giga\hertz} and \SI{450}{\giga\hertz} for the use of land mobile and fixed services, paving the way of deploying THz commutations in 6G \cite{Ref_kuerner2020impact}.

Despite its high potential, mmWave and THz communications suffer from severe propagation losses raised from high free-space path loss, atmospheric gaseous absorption, and weather attenuation \cite{Ref_siles2015atmospheric}, leading to short transmission distances. In addition, high-frequency transmission is more susceptible to blockage. Hence, antenna arrays, typically using hybrid analog-digital beamforming \cite{Ref_jiang2022initial_ICC}, are required at the base station (BS) and/or the mobile terminal to achieve sufficient beamforming gains for coverage extension. 
Recently, a disruptive technique called intelligent reflecting surface (IRS) a.k.a reconfigurable intelligent surface has received much attention. It can significantly boost the spectral and power efficiency of wireless communications by smartly re-configuring  propagation environment. Specifically, IRS is a planar surface comprising a large number of small, passive, cheap reflecting elements, each being able to independently induce a phase shift and/or amplitude attenuation (collectively termed as reflection coefficient) to the incident signal, thereby collaboratively achieving high passive beamforming gain on the order of $N^2$, where $N$ is the number of reflecting elements.

Therefore, it is worth exploiting the synergy of hybrid beamforming and IRS to further improve the coverage of mmWave and THz communications in the upcoming 6G networks, e.g., \cite{Ref_di2020hybrid, Ref_di2021terahertz}.  
In this context, this paper proposes a novel scheme, coined dual-beam IRS (DB-IRS), for IRS-enhanced systems operating in high-frequency bands. Making use of hybrid beamforming, the BS generates a pair of beams towards the IRS and user equipment (UE), respectively. As a result, the optimal reflecting phases are directly calculated from the estimated channel state information (CSI), independently with the active beamforming. Hence, the optimization of passive and active beamforming is decoupled, resulting in a simplified system design. In contrast to \cite{Ref_wu2019intelligent}, where the passive and active beamforming is jointly optimized,  the high computational burden and high latency due to iterative optimization can be avoided. This paper provides numerical results to corroborate the optimized coverage in terms of the cell-edge and cell-center performance, and compare it with several optimal and sub-optimal schemes. 

The remainder of this paper is organized as follows: the next section introduces the system model and optimization formulation. Section III explains the principle of dual-beam IRS and its optimization design. Numerical results are demonstrated in Section IV, followed by the conclusions.

\section{Multi-Antenna IRS Transmission}
\subsection{System Model}
As \cite{Ref_wu2019intelligent}, we consider a single-cell multi-input single-output (MISO) communications system comprising an $N_b$-antenna BS, a single-antenna UE, and an IRS with $N$ reflecting elements. The IRS is equipped with a smart controller, which can dynamically adjust the phase shift of each reflecting element in terms of the instantaneous CSI acquired through periodic estimation. In particular, several channel estimation methods for IRS-aided systems have been proposed to obtain accurate CSI, such as \cite{Ref_you2020channel, Ref_wang2020channel, Ref_zheng2020intelligent}, which is beyond the scope of this paper. To characterize the theoretical performance, we assume the CSI of all involved channels is perfectly known and follows quasi-static frequency-flat fading. A wideband channel suffers from frequency selectivity, but it can be transformed into a set of narrowband channels through OFDM \cite{Ref_jiang2016ofdm}, making the assumption of flat fading reasonable. Due to the high path loss, the signals that are reflected by the IRS twice or more are negligible. Since the IRS is passive, we adopt time-division duplexing (TDD) operation and assume perfect channel reciprocity. Although we focus on the downlink transmission from the BS to the UE, the technique is also applicable to the uplink.

The BS applies linear  beamforming denoted by a transmit vector $\mathbf{w}\in \mathbb{C}^{N_b\times 1}$, with $\|\mathbf{w}\|^2\leqslant 1$, where $\|\cdot\|$ represents the Euclidean norm of a complex vector. Then, the discrete-time baseband signal received by the user is expressed as
\begin{equation} \label{eqn_systemModel}
    r=\sqrt{P_t}\Biggl( \sum_{n=1}^N g_n c_n \mathbf{h}_n^T + \mathbf{h}_d^T \Biggr) \mathbf{w} s + n,
\end{equation}
where $s$ denotes a normalized transmitted symbol satisfying $\mathbb{E}\left[|s|^2\right]=1$, $P_t$ expresses the transmitted power of the BS, $n$ is additive white Gaussian noise (AWGN) with zero mean and variance $\sigma_n^2$, i.e., $z\sim \mathcal{CN}(0,\sigma_n^2)$, $\mathbf{h}_d=[h_{d,1},h_{d,2},\ldots,h_{d,N_b}]^T\in \mathbb{C}^{N_b\times 1}$ is the channel vector from $N_b$ base-station antennas to the user,  $\mathbf{h}_{n}=[h_{n,1},h_{n,2},\ldots,h_{n,N_b}]^T\in \mathbb{C}^{N_b\times 1}$ is the channel vector from $N_b$ base-station antennas to the $n^{th}$ reflecting element, $g_{n}$ denotes the channel coefficient from the $n^{th}$ reflecting element to the user. 

We write $c_n=\beta_n e^{j\phi_n}$ to denote the reflection coefficient of the $n^{th}$ IRS element, with the induced phase shift $\phi_n\in [0,2\pi)$ and the amplitude attenuation $\beta_n\in [0,1]$. As revealed by \cite{Ref_wu2019intelligent}, the optimal value of reflecting attenuation is $\beta_n=1$, $\forall n=1,2,\ldots,N$ to maximize the received power and simplify the hardware implementation. Let $\mathbf{g}=[g_1,g_2,\ldots,g_N]^T$,  $\boldsymbol{\Theta}=\mathrm{diag}\{e^{j\phi_1},\ldots,e^{j\phi_N}\}$, and  $\mathbf{H}\in \mathbb{C}^{N\times N_b}$ to denote the channel matrix from the BS to the IRS, where the $n^{th}$ row of $\mathbf{H}$ equals to $\mathbf{h}_n^T$, \eqref{eqn_systemModel} can be rewritten in matrix form as
\begin{equation}
    r= \sqrt{P_t}\Bigl(\mathbf{g}^T \boldsymbol{\Theta} \mathbf{H} +\mathbf{h}_d^T\Bigr)\mathbf{w} s +n.
\end{equation}

\subsection{Problem Formulation}
By jointly designing the active beamforming $\mathbf{w}$ and passive beamforming $\boldsymbol{\Theta}$, the achievable spectral efficiency of IRS
\begin{equation}
    R=\log\left(1+\frac{P_t \Bigl|\bigl(\mathbf{g}^T \boldsymbol{\Theta} \mathbf{H} +\mathbf{h}_d^T\bigr)\mathbf{w}\Bigr|^2 }{\sigma_n^2} \right)
\end{equation}
can be maximized, resulting in the following optimization problem
\begin{equation}
\begin{aligned} \label{eqnIRS:optimizationMRTvector}
\max_{\boldsymbol{\Theta},\:\mathbf{w}}\quad &  \biggl|\Bigl(\mathbf{g}^T \boldsymbol{\Theta} \mathbf{H} +\mathbf{h}_d^T\Bigr)\mathbf{w}\biggr|^2\\
\textrm{s.t.} \quad & \|\mathbf{w}\|^2\leqslant 1\\
  \quad & \phi_n\in [0,2\pi), \: \forall n=1,2,\ldots,N.
\end{aligned}
\end{equation}
Unfortunately, it is a non-convex optimization problem because its objective function is not jointly concave with respect to $\boldsymbol{\Theta}$ and $\mathbf{w}$. The \textit{coupling} between the active and passive beamforming variables is the major challenge in their joint optimization. Although two solutions, i.e., semi-definite relaxation (SDR) and alternating optimization (AO), have been proposed in \cite{Ref_wu2019intelligent}, their iterative optimization process raises a high computational burden and high latency, which are not favorable in a practical system. Hence, it's worth exploring a simple scheme for an IRS-aided MISO system.

\section{Dual-Beam Intelligent Reflecting Surface}
In this context, we propose a dual-beam transmission scheme, where a pair of beams are formed over a hybrid analog-digital beamformer.  Steering one beam towards the user, while pointing the other beam to the IRS, the joint passive and active optimization can be decoupled and the system design is therefore simplified.

\subsection{Dual Beams over Hybrid Beamforming}
The propagation of mmWave and THz signals relies heavily on high beamforming gain to extend coverage. There are three major implementation structures: digital, analog, and hybrid digital-analog. Implementing digital beamforming in a mmWave or THz transceiver equipped with a large-scale array needs a large number of radio frequency (RF) components, leading to high hardware cost and power consumption. This constraint has driven the application of analog beamforming, using only a single RF chain. It has been adopted as the \textit{de-facto} approach for indoor mmWave systems. However, it only supports single-stream transmission and suffers from analog hardware impairments, e.g., phase noise \cite{Ref_jiang2021impactcellfree}.

As a consequence, hybrid beamforming \cite{Ref_zhang2019hybrid} becomes attractive as an efficient approach to support multi-stream transmission with only a few RF chains. It achieves spectral efficiency comparable to digital beamforming but the hardware complexity and cost is much lower. Hence, it is promising for mmWave and THz communications in 6G systems.
Hybrid beamforming can be implemented with different forms of circuit networks, resulting in two basic structures:
\textit{Fully-Connected Hybrid Beamforming}, where each RF chain is connected to all antennas via an analog network, and \textit{Partially-Connected Hybrid Beamforming}, where each RF chain is only connected to a sub-array consisting of a subset of antenna elements. Without losing generality, we use the partially-connected structure with two branches hereinafter to present and analyze the proposed scheme but the results are applicable to the fully-connected structure.

\begin{figure}[!bpht]
\centering
\includegraphics[width=0.42\textwidth]{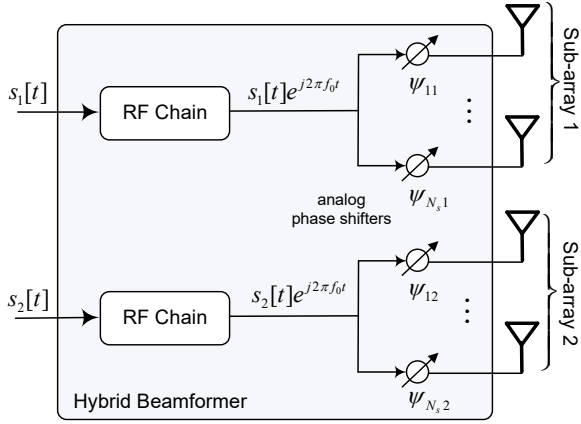}
\caption{Block diagrams of the partially-connected hybrid beamforming. }
\label{Diagram_hybridBF}
\end{figure}

Mathematically, the baseband input signals are denoted by $s_m[t]$, $m=1,2$, as shown in \figurename \ref{Diagram_hybridBF}. After the digital-to-analog conversion and up-conversion, the $m^{th}$ RF chain feeds
\begin{equation}
    \Re\Bigr [s_m[t]e^{j2\pi f_0 t}\Bigl]
\end{equation}
into the analog network, where $\Re [\cdot]$ denotes the real part of a complex number, and $f_0$ stands for the carrier frequency.  In this two-branch partially-connected structure, the antenna array is divided into two sub-arrays, and each antenna is connected to one RF chain. Thus, each sub-array contains $N_s=N_b/2$ elements (suppose $N_s$ is an integer).  We write $\psi_{nm}$,  $n=1,\ldots, N_s$ to denote the weighted phase on the $n^{th}$ antenna of the $m^{th}$ sub-array, and thus the vector of transmit signals on the $m^{th}$ sub-array is given by
\begin{align} \label{eqn:mmWave:TxSingalPerSubarray}
    &\textbf{s}_m(t) = \\ \nonumber
    &\biggl[\Re\left [s_m[t]e^{j2\pi f_0t}e^{j\psi_{1m}}\right],\ldots,\Re\left [s_m[t]e^{j2\pi f_0t}e^{j\psi_{N_sm}}\right] \biggr]^T.
\end{align}

Radiating a plane wave into a homogeneous media in the direction indicated by the angle of departure $\theta$, the time difference between a typical element $n$  and the reference point of the $m^{th}$ sub-array is denoted by $\tau_{nm}(\theta)$.
As derived in \cite{Ref_jiang2022initial_ICC}, the received passband signal is
\begin{equation} \nonumber \label{eqn:mmWave:BeamPatternPahyBF}
    r(t) =   \Re\left [\left( \sum_{m=1}^M h_m(t) s_m[t] \textbf{a}_m^T(\theta) \textbf{w}_m  \right)e^{j2\pi  f_0t}\right] +n(t),
\end{equation}
where $h_m(t)$ expresses the channel response between the $m^{th}$ sub-array and UE, $n(t)$ is additive noise, $\textbf{w}_m$ denotes the steering vector of sub-array $m$, i.e.,
\begin{equation}
    \textbf{a}_m(\theta) = \left[e^{-j2\pi f_0\tau_{1m}(\theta)},\ldots,e^{-j2\pi f_0\tau_{N_sm}(\theta)}   \right]^T,
\end{equation}
and its weighting vector (corresponding to analog phase shifts)
\begin{equation}
    \textbf{w}_m = \left[e^{j\psi_{1m}}, e^{j\psi_{2m}},\ldots, e^{j\psi_{N_sm}}   \right]^T.
\end{equation}

Therefore, the beam pattern generated by the $m^{th}$ sub-array can be expressed as
\begin{equation} \label{eqn:ICC:beampattern}
   B_m(\theta)= \textbf{a}_m^T(\theta)\textbf{w}_m=\sum_{n=1}^{N_s}  e^{-j2\pi f_0\tau_{nm}(\theta)}e^{j\psi_{nm}}.
\end{equation}
Then, a pair of beams with patterns $B_1(\theta)$ and $B_2(\theta)$ over the hybrid beamformer are obtained. 
After the down-conversion and sampling at the receiver, the baseband received signal can be simplified into (neglecting the time index for simplicity)
\begin{equation} \label{eqn:ICC:BBRxSignal}
    r =   h_1 B_1(\theta_1) s_1 + h_2 B_2(\theta_2) s_2 + n, \end{equation}
where $h_1$ and $h_1$ denote the channel coefficients from the reference antenna of the first and second sub-arrays to the UE,  and $\theta_1$ and $\theta_2$ stand for the departure-of-angles (DOA) of the transmitted signals over sub-arrays 1 and 2, respectively.

\begin{figure}[!bpht]
\centering
\includegraphics[width=0.45\textwidth]{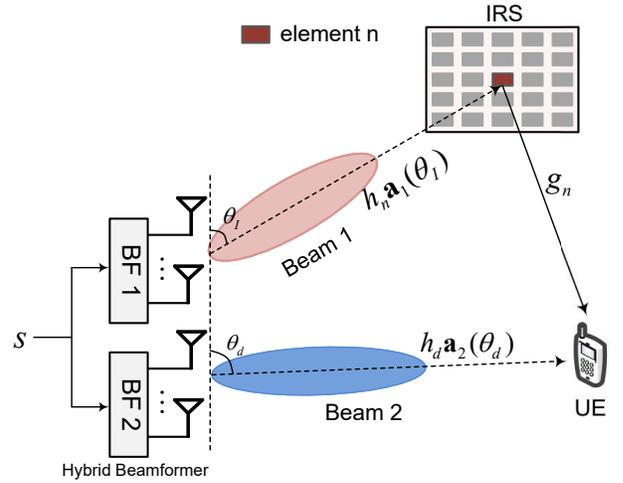}
\caption{Illustration of the dual-beam IRS using hybrid beamforming. }
\label{Diagram_DBIRS}
\end{figure}

\subsection{Dual-Beam IRS}
The fundamental of DB-IRS is to steer one beam towards the user, like a conventional beamforming system, while the other beam is used to focus energy on the IRS, as demonstrated in \figurename \ref{Diagram_DBIRS}. Since both the locations of the IRS and BS are deliberately selected and fixed, there is usually no blockage in-between and the DOA information $\theta_I$ is deterministic. The beam towards the IRS is fixed, which substantially simplifies the system implementation.
The inter-antenna spacing in hybrid beamforming is small, typically half wavelength $d=\lambda/2$. As a result, the antennas are highly correlated.
Without losing generality, we assume the first sub-array serves the IRS, and the second sub-array towards the UE. The channel vector from the first sub-array to the $n^{th}$ reflecting element can be expressed as $\mathbf{h}_n=h_n\textbf{a}_1(\theta_I)\in \mathbb{C}^{N_s\times 1}$, as derived in \cite{Ref_yang2013random}, where $h_n$ denotes the channel response from the reference antenna of this sub-array to the $n^{th}$ reflecting element. Likewise, the channel vector from the second sub-array to the UE is given by $\mathbf{h}_d=h_d\textbf{a}_2(\theta_d)$, where $h_d$ denotes the channel response from the reference antenna of the second sub-array to the UE, and $\theta_d$ is the DOA of the user.
We write $\mathbf{w}_1$ and  $\mathbf{w}_2$ to denote the weighting vectors of two sub-arrays, respectively. Meanwhile, we assume two sub-arrays have a common transmitted symbol $s$ with $\mathbb{E}\left[|s|^2\right]=1$.
Referring to \eqref{eqn_systemModel}, the observation at the user in a DB-IRS system can be given by
\begin{align} \nonumber
    r &= \sqrt{P_t}\left(\sum_{n=1}^N  g_n c_n \mathbf{h}_n^T \mathbf{w}_1 + \mathbf{h}_d^T  \mathbf{w}_2\right)s+n\\ \nonumber
    &= \sqrt{P_t}\left(\sum_{n=1}^N  g_n c_n h_n \textbf{a}_1^T(\theta_I)  \mathbf{w}_1 + h_d \textbf{a}_2^T(\theta_d)  \mathbf{w}_2\right)s+n\\
    &= \sqrt{P_t}\left(\sum_{n=1}^N g_n c_n h_n  B_r(\theta_I)  + h_d B_d(\theta_d)\right) s +n,
\end{align}
with the beam patterns corresponding to the IRS and UE $B_r(\theta)=\textbf{a}_1^T(\theta)\mathbf{w}_1$ and $B_d(\theta)=\textbf{a}_2^T(\theta)\mathbf{w}_2$, respectively.

\subsection{Optimization Design}

Accordingly, the received SNR at the user is given by
\begin{equation} \label{eqnDBIRSsnr}
    \gamma = \frac{P_t \Bigl|\sum_{n=1}^N g_n c_n h_n  B_r(\theta_I)  + h_d B_d(\theta_d)\Bigr|^2 }{\sigma_n^2}.
\end{equation}
The aim of the optimization design is to maximize the spectral efficiency $  R=\log\left(1+\gamma \right)$ through selecting the optimal weighting vectors and reflecting coefficients, resulting in the following optimization problem
\begin{equation} \label{IRSoptimizationPrlbme}
\begin{aligned}
\max_{\boldsymbol{\Theta},\mathbf{w}_m}\quad &  \biggl|\sum_{n=1}^N g_n c_n h_n  B_r(\theta_I)  + h_d B_d(\theta_d)\biggr|^2\\
\textrm{s.t.} \quad & \|\mathbf{w}_m\|^2\leqslant 1,\:m=1,2,\\
  \quad & \phi_n\in [0,2\pi), \: \forall n=1,2,\ldots,N.
\end{aligned}
\end{equation}
Mostly importantly, $\boldsymbol{\Theta}$ and $\mathbf{w}_m$, $m=1,2$ in this optimization problem is decoupled thanks to the dual-beam approach. Thus, $\boldsymbol{\Theta}$ and $\mathbf{w}_m$, $m=1,2$ can be optimized independently, in contrast to the joint optimization in \eqref{eqnIRS:optimizationMRTvector}. In addition, the determination of $\mathbf{w}_1$ and $\mathbf{w}_2$ is also independent.

To get the optimal beam towards the IRS, we need to solve
\begin{equation}
\begin{aligned}
\max_{\mathbf{w}_1} \quad & \Bigl|\textbf{a}_1^T(\theta_I)  \mathbf{w}_1\Bigl|^2   \\
\textrm{s.t.} \quad & \|\mathbf{w}_1\|^2\leqslant 1.
\end{aligned}
\end{equation}
What we need to know is the DOA information, which can be estimated through classical algorithms efficiently, such as MUSIC and ESPRIT \cite{Ref_MUSIC}. Given the estimated $\theta_I$, the optimal weighting vector is $\mathbf{w}_1=\sqrt{1/N_s}\textbf{a}_1^*(\theta_I)$, resulting in
\begin{equation} \label{optimalWeight1}
    B_r(\theta_I)=\sqrt{\frac{1}{N_s}} \textbf{a}_1^T(\theta_I)  \textbf{a}_1^*(\theta_I)=\sqrt{N_s}.
\end{equation}
Since the BS and IRS are fixed, $\theta_d$ is easier to obtain and keeps constant in a long-term basis.
Similarly, the direct beam gain $B_d(\theta_d)=\sqrt{N_s}$ if the weighting vector is set to
\begin{equation} \label{optimalWeight2}
    \mathbf{w}_2=\sqrt{\frac{1}{N_s}}\textbf{a}_2^*(\theta_d).
\end{equation} For simplicity, we assume that the sidelobes of two beam patterns in the other's direction are negligible, i.e., $|B_d(\theta_I)|=0$ and $|B_r(\theta_d)|=0$.

Afterwards, the optimization problem in \eqref{IRSoptimizationPrlbme} is reduced to
\begin{equation}
\begin{aligned}
\max_{\boldsymbol{\Theta}}\quad & N_s \biggl|\sum_{n=1}^N g_n c_n h_n   + h_d \biggr|^2\\
\textrm{s.t.}  \quad & \phi_n\in [0,2\pi), \: \forall n=1,2,\ldots,N,
\end{aligned}
\end{equation}
which is equivalent to an IRS system with a single-antenna BS, as given in Eq. (18) of \cite{Ref_wu2021intelligent}.
Therefore, the optimal phase shift for reflecting element $n$ equals to
\begin{equation} \label{IRSeqn:optimalphase}
    \phi_n^\star = \mod \Bigl[\phi_{d}-(\phi_{h,n}+\phi_{g,n}), 2\pi   \Bigr],
\end{equation}
where $\phi_{h,n}$, $\phi_{g,n}$, and $\phi_{d}$ stand for the phases of $h_n$, $g_n$, and $h_d$, respectively, and $\mod[\cdot]$ is the modulo operation.
Given the optimal beams in \eqref{optimalWeight1} and \eqref{optimalWeight2} and the optimal phase shifts in \eqref{IRSeqn:optimalphase}, the received SNR of the dual-beam IRS system, as given by \eqref{eqnDBIRSsnr},  is maximized, equaling to
\begin{equation}
    \gamma_{max} = \frac{N_sP_t  \Bigl|\sum_{n=1}^N |g_n|| h_n|   + |h_d| \Bigr|^2 }{\sigma_n^2},
\end{equation}
implying an active beamforming gain of $N_s=N_b/2$ and a passive beamforming gain of $N^2$.

Compared to the joint active and passive beamforming optimization in \cite{Ref_wu2019intelligent}, which needs an iterative optimization process, the calculation of dual-beam IRS parameters is substantially simple. In a nutshell, a DB-IRS system only needs to estimate the DOA of the user, and then steers a beam towards it. The beam towards the IRS is fixed, which needs an operation in a long-term basis since the locations of the IRS and BS are fixed and known. Meanwhile, the optimal phase shifts for the IRS are obtained in terms of the acquired CSI directly, as \eqref{IRSeqn:optimalphase}. The proposed DB-IRS algorithm is depicted in \algorithmcfname \ref{alg:IRS001}.

\SetKwComment{Comment}{/* }{ */}
\RestyleAlgo{ruled}
\begin{algorithm}
\caption{Dual-Beam IRS} \label{alg:IRS001}
\SetKwInOut{Input}{input}\SetKwInOut{Output}{output} \SetKwInput{kwInit}{Initialization}
\Input{$\theta_I$}
\kwInit {$\mathbf{w}_1=\sqrt{1/N_s}\textbf{a}_1^*(\theta_I)$}
\ForEach{Channel coherence time}{
  Estimate $h_n$,$g_n$,$\forall n$, $h_d$ \cite{Ref_you2020channel, Ref_wang2020channel, Ref_zheng2020intelligent}\;
  Compute $\phi_n^\star = \mod \Bigl[\phi_{d}-(\phi_{h,n}+\phi_{g,n}), 2\pi   \Bigr]$\;
  Adjust the IRS using $\phi_n^\star$\;
  Estimate $\theta_d$ \cite{Ref_MUSIC}\;
  Compute $\mathbf{w}_2=\sqrt{1/N_s}\textbf{a}_2^*(\theta_d)$\;
  \For{t=1:T}
 {Transmit $\mathbf{w}_1 s_t$ over sub-array 1\;
  Transmit $\mathbf{w}_2 s_t$ over sub-array 2\;}
  }
\end{algorithm}

\section{Numerical results}
This part provides some typical numerical examples to corroborate the effectiveness of the proposed DB-IRS technique and compare its performance with the related schemes. For the sake of comparison, we use the same simulation setup of a three-dimensional coordinate system given by \cite{Ref_wu2019intelligent}, as depicted by \figurename 2 of that article. Hybrid beamforming with two RF branches and two uniform linear sub-arrays is employed at the base station, given the inter-antenna spacing of half wavelength.
The distance-dependent path loss is computed by
\begin{equation}
    L(d)=\frac{L_0}{d_0^{-\alpha}},
\end{equation}
where $L_0$ is the path loss at the reference distance of \SI{1}{\meter}, $d_0$ stands for the propagation distance, and $\alpha$ means the path loss exponent.
In addition, taking into account the shadowing, and the special condition in mmWave and THz communications, i.e., atmospheric gaseous absorption and weather attenuation \cite{Ref_siles2015atmospheric}, we assume an extra loss of \SI{10}{\decibel} for both the $\mathbb{B}-\mathbb{U}$ and  $\mathbb{I}-\mathbb{U}$ links, while the $\mathbb{B}-\mathbb{I}$ link should be light-of-sight without any blockage due to the favourable locations of the BS and IRS. In this regard, the Rician channel model can provide small-scale fading for all involved channels, i.e.,
\begin{equation}
    \mathbf{h}=\sqrt{\frac{K}{K+1}}\mathbf{h}_{LOS} + \sqrt{\frac{1}{K+1}}\mathbf{h}_{NLOS},
\end{equation}
with the Rician factor $K$, the LOS component $\mathbf{h}_{LOS}$, and the multipath (Rayleigh) component $\mathbf{h}_{NLOS}$.

To align with \cite{Ref_wu2019intelligent} at most, we follow the setup where the BS and IRS are apart from $d_{BI}=51\si{\meter}$ and the UE lies on a horizontal line with the vertical distance of $d_v=2\si{\meter}$. We write the horizontal distance between the base station and user by $d$. Accordingly, the $\mathbb{B}-\mathbb{U}$ and  $\mathbb{I}-\mathbb{U}$ distances are computed by $d_{BU}=\sqrt{d^2+d_v^2}$ and $d_{IU}=\sqrt{(d_{BI}-d)^2+d_v^2}$, respectively. Then, we can investigate the received SNR of the user in terms of the value of $d$. Other simulation parameters are provided as follows: $N=200$, $N_b=16$, $L_0=\SI{-30}{\decibel}$, $P_t=5\mathrm{dBm}$, $\sigma_n^2=-80\mathrm{dBm}$, the Rician factor and path loss component for the $\mathbb{B}-\mathbb{U}$, $\mathbb{B}-\mathbb{I}$, and $\mathbb{I}-\mathbb{U}$ links are  $\{K=0$, $\alpha=3\}$, $\{K=+\infty$, $\alpha=2\}$, and $\{K=0$, $\alpha=3\}$, respectively.
\begin{figure}[!bpht]
\centering
\includegraphics[width=0.445\textwidth]{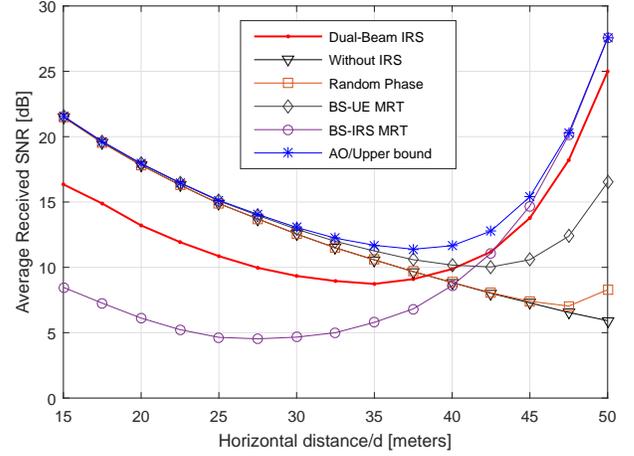}
\caption{Results of the received SNR versus the horizontal distance. }
\label{Diagram_Results}
\end{figure}

\figurename \ref{Diagram_Results} compares the receive SNR of the following schemes: 1) Alternating optimization  for the joint active and passive beamforming proposed in \cite{Ref_wu2019intelligent}. In our simulations, the number of iterations is set to three, which is enough for well convergence. Since AO achieves the optimal results as the same as the SDR scheme or the upper performance bound, the other two are not shown in the figure for simplicity. 2) BS-UE maximum-ratio transmission (MRT), which sets $\mathbf{w}^\star=\mathbf{h}_d^*/\|\mathbf{h}_d\|$ to achieve matched filtering in terms of the $\mathbb{B}-\mathbb{U}$ direct link; 3) BS-IRS MRT that sets $\mathbf{w}^\star=\mathbf{h}_n^*/\|\mathbf{h}_n\|$ to achieve matched filtering based on the $\mathbb{B}-\mathbb{I}$ link, while the optimal reflecting phases are computed according to (28) in \cite{Ref_wu2019intelligent}. Since this link is LOS, or a rank-one channel, $\mathbf{h}_n$ can be any row of $\mathbf{H}$. 4) Random phase shifts: we set $\theta_n$, $\forall n$ randomly and then perform MRT at the BS based on the combined channel, i.e., $\mathbf{w}^\star = \frac{(\mathbf{g}^T \boldsymbol{\Theta} \mathbf{H} +\mathbf{h}_d^T)^*}{\|\mathbf{g}^T \boldsymbol{\Theta} \mathbf{H} +\mathbf{h}_d^T\|}$. 
5) In addition, we observe a benchmark scheme in a MISO system without the aid of IRS by setting $\mathbf{w}^\star=\mathbf{h}_d^*/\|\mathbf{h}_d\|$.

In a conventional system without IRS, the cell-edge user suffers from low SNR due to severe propagation loss. The cell-edge problem is alleviated by deploying an IRS, as shown in \figurename \ref{Diagram_Results}, where the IRS-aided wireless networks exhibit comparable cell-edge performance as the cell center. This is because the user farther away from the BS is closer to the IRS and thus it is able to get strong reflected signals. It implies that the system coverage can be effectively extended by deploying passive IRS in addition to a BS or active relay. Specially, the BS-UE MRT scheme performs near-optimally when the UE is at the cell center, whereas suffering from a considerable SNR loss at the cell edge. That is because the received signal is dominated by the direct link at the cell center whereas the reflected link is dominant at the cell edge. Moreover, it can be observed that BS-IRS MRT behaves oppositely as the UE moves away from the BS towards the IRS. It also reveals the significance of accurate phase shifts since random phase shifts overwhelm the potential of IRS. The proposed DB-IRS scheme can achieve a good balance at the cell center and cell edge. Specifically, its cell-edge performance approaches to the optimal result but the implementation complexity is reduced.

\section{Conclusions}
This paper proposed a novel transmission scheme for intelligent reflecting surface-enhanced mmWave and THz systems. Making use of hybrid beamforming, the BS generates a pair of beams towards the IRS and UE, respectively. The major benefit of dual-beam IRS is that the optimization of passive and active beamforming can be decoupled, resulting in a simplified system design. To be specific, the optimal reflecting phases can be calculated from the CSI independently, without iterative optimization. Meanwhile, the active beamforming is only based on the DOA information, which is simple to acquire. Its performance is well balanced between the cell edge and cell center, with a moderate loss compared with the performance bound, but it still remarkably outperforms the other sub-optimal schemes.





%

\bibliographystyle{IEEEtran}
\bibliography{IEEEabrv,Ref_VTC2022}

\begin{thebibliography}{10}
\providecommand{\url}[1]{#1}
\csname url@samestyle\endcsname
\providecommand{\newblock}{\relax}
\providecommand{\bibinfo}[2]{#2}
\providecommand{\BIBentrySTDinterwordspacing}{\spaceskip=0pt\relax}
\providecommand{\BIBentryALTinterwordstretchfactor}{4}
\providecommand{\BIBentryALTinterwordspacing}{\spaceskip=\fontdimen2\font plus
\BIBentryALTinterwordstretchfactor\fontdimen3\font minus
  \fontdimen4\font\relax}
\providecommand{\BIBforeignlanguage}[2]{{%
\expandafter\ifx\csname l@#1\endcsname\relax
\typeout{** WARNING: IEEEtran.bst: No hyphenation pattern has been}%
\typeout{** loaded for the language `#1'. Using the pattern for}%
\typeout{** the default language instead.}%
\else
\language=\csname l@#1\endcsname
\fi
#2}}
\providecommand{\BIBdecl}{\relax}
\BIBdecl

\bibitem{Ref_jiang2021kickoff}
W.~Jiang and H.~D. Schotten, ``The kick-off of {6G} research worldwide: An
  overview,'' in \emph{Proc. 2021 Seventh IEEE Int. Conf. on Comput. and
  Commun. (ICCC)}, Chengdu, China, Dec. 2021.

\bibitem{Ref_jiang2021road}
W.~Jiang \emph{et~al.}, ``The road towards {6G}: A comprehensive survey,''
  \emph{IEEE Open J. Commun. Society}, vol.~2, pp. 334--366, Feb. 2021.

\bibitem{Ref_jiang2022initial}
W.~Jiang and H.~D. Schotten, ``Initial beamforming for millimeter-wave and
  terahertz communications in {6G} mobile systems,'' in \emph{Proc. 2022 IEEE
  Wireless Commun. and Netw. Conf. (WCNC)}, Austin, USA, Apr. 2022.

\bibitem{Ref_kuerner2020impact}
T.~Kuerner and A.~Hirata, ``On the impact of the results of {WRC 2019 on THz}
  communications,'' in \emph{Proc. of 2020 Third Intl. Workshop on Mobile
  Terahertz Syst. (IWMTS)}, Essen, Germany, Jul. 2020, pp. 1--3.

\bibitem{Ref_siles2015atmospheric}
G.~A. Siles, J.~M. Riera, and P.~G. del Pino, ``Atmospheric attenuation in
  wireless communication systems at millimeter and {THz} frequencies,''
  \emph{{IEEE} Antennas Propag. Mag.}, vol.~57, no.~1, pp. 48 -- 61, Feb. 2015.

\bibitem{Ref_jiang2022initial_ICC}
W.~Jiang and H.~D. Schotten, ``Initial access for millimeter-wave and terahertz
  communications with hybrid beamforming,'' in \emph{Proc. 2022 IEEE Int.
  Commun. Conf. (ICC)}, Seoul, South Korea, May 2022.

\bibitem{Ref_di2020hybrid}
B.~Di \emph{et~al.}, ``Hybrid beamforming for reconfigurable intelligent
  surface based multi-user communications: Achievable rates with limited
  discrete phase shifts,'' \emph{{IEEE} J. Sel. Areas Commun.}, vol.~38, no.~8,
  pp. 1809 -- 1822, Aug. 2020.

\bibitem{Ref_di2021terahertz}
B.~Ning \emph{et~al.}, ``Terahertz multi-user massive {MIMO} with intelligent
  reflecting surface: Beam training and hybrid beamforming,'' \emph{{IEEE}
  Trans. Veh. Technol.}, vol.~70, no.~2, pp. 1376 -- 1393, Feb. 2021.

\bibitem{Ref_wu2019intelligent}
Q.~Wu and R.~Zhang, ``Intelligent reflecting surface enhanced wireless network
  via joint active and passive beamforming,'' \emph{{IEEE} Trans. Wireless
  Commun.}, vol.~18, no.~11, pp. 5394 -- 5409, Nov. 2019.

\bibitem{Ref_you2020channel}
C.~You, B.~Zheng, and R.~Zhang, ``Channel estimation and passive beamforming
  for intelligent reflecting surface: Discrete phase shift and progressive
  refinement,'' \emph{{IEEE} J. Sel. Areas Commun.}, vol.~38, no.~11, pp. 2604
  -- 2620, Nov. 2020.

\bibitem{Ref_wang2020channel}
Z.~Wang, L.~Liu, and S.~Cui, ``Channel estimation for intelligent reflecting
  surface assisted multiuser communications: Framework, algorithms, and
  analysis,'' \emph{{IEEE} Trans. Wireless Commun.}, vol.~19, no.~10, pp. 6607
  -- 6620, Oct. 2020.

\bibitem{Ref_zheng2020intelligent}
B.~Zheng and R.~Zhang, ``Intelligent reflecting surface-enhanced {OFDM}:
  Channel estimation and reflection optimization,'' \emph{IEEE Wireless Commun.
  Lett.}, vol.~9, no.~4, pp. 518 -- 522, Apr. 2020.

\bibitem{Ref_jiang2016ofdm}
W.~Jiang and T.~Kaiser, ``From {OFDM} to {FBMC}: Principles and
  {Comparisons},'' in \emph{Signal Processing for 5G: Algorithms and
  Implementations}, F.~L. Luo and C.~Zhang, Eds.\hskip 1em plus 0.5em minus
  0.4em\relax United Kindom: John Wiley\&Sons and IEEE Press, 2016, ch.~3.

\bibitem{Ref_jiang2021impactcellfree}
W.~Jiang and H.~Schotten, ``Impact of channel aging on zero-forcing precoding
  in cell-free massive {MIMO} systems,'' \emph{{IEEE} Commun. Lett.}, vol.~25,
  no.~9, pp. 3114 -- 3118, Sep. 2021.

\bibitem{Ref_zhang2019hybrid}
J.~Zhang \emph{et~al.}, ``Hybrid beamforming for {5G} and beyond
  millimeter-wave systems: A holistic view,'' \emph{IEEE Open J. Commun.
  Society}, vol.~1, pp. 77 -- 91, 12 2019.

\bibitem{Ref_yang2013random}
X.~Yang, W.~Jiang, and B.~Vucetic, ``A random beamforming technique for
  omnidirectional coverage in multiple-antenna systems,'' \emph{{IEEE} Trans.
  Veh. Technol.}, vol.~62, no.~3, pp. 1420 -- 1425, Mar. 2013.

\bibitem{Ref_MUSIC}
W.~Gardner, ``Simplification of {MUSIC} and {ESPRIT} by exploitation of
  cyclostationarity,'' \emph{Proc. of the IEEE}, vol.~76, no.~7, pp. 845--847,
  Jul. 1988.

\bibitem{Ref_wu2021intelligent}
Q.~Wu, S.~Zhang, B.~Zheng, C.~You, and R.~Zhang, ``Intelligent reflecting
  surface-aided wireless communications: A tutorial,'' \emph{{IEEE} Trans.
  Commun.}, vol.~69, no.~5, pp. 3313 -- 3351, May 2021.

\end{thebibliography}

\end{document}